\documentclass[a4paper,showpacs,prb,twocolumn]{revtex4}
\usepackage{amsfonts}
\usepackage{amsmath}
\usepackage{amssymb}
\usepackage{graphicx}

\begin{document}

\title{Light propagation in finite and infinite photonic crystals: The
recursive Green's function technique. }
\author{A. I. Rahachou and I. V. Zozoulenko}
\affiliation{Department of Science and Technology, Link\"{o}ping University 601 74, Norrk%
\"{o}ping, Sweden}
\date{\today }

\begin{abstract}
We report a new computational method based on the recursive
Green's function technique for calculation of light propagation in
photonic crystal structures. The advantage of this method in
comparison to the conventional finite-difference time domain
(FDTD) technique is that it computes Green's function of the
photonic structure \textit{recursively} by adding slice by slice
on the basis of Dyson's equation. This eliminates the need for
storage of the wave function in the whole structure, which
obviously strongly relaxes the memory requirements and enhances
the computational speed. The second advantage of this method is
that it can easily account for the infinite extension of the
structure both into air and into the space occupied by the
photonic crystal by making use of the so-called ``surface Green's
functions". This eliminates the spurious solutions (often present
in the conventional FDTD methods) related to e.g. waves reflected
from the boundaries defining the computational domain. The
developed method has been applied to study scattering and
propagation of the electromagnetic waves in the photonic band-gap
structures including cavities and waveguides. A particular
attention has been paid to surface modes residing on a termination
of a semi-infinite photonic crystal. We demonstrate that coupling
of the surface states with incoming radiation may result in
enhanced intensity of an electromagnetic field on the surface and very high $%
Q$ factor of the surface state. This effect can be employed as an
operational principle for surface-mode lasers and sensors.
\end{abstract}

\maketitle

\section{\protect\bigskip Introduction}

Optical microcavities and photonic crystals (PC) have received
increased attention in recent years because of the promising
prospects of applications in a future generation of optical
communication networks \cite{JP,Sakoda}. Examples of successfully
demonstrated devices include lasers, light emitting
diodes, waveguides, add-drop filters, delay lines, and many other \cite%
{review}.

By far the most popular method for the theoretical description of light
propagation in these systems is the finite-difference time-domain method
(FDTD) introduced by Yee \cite{Yee}. The success of the FDTD method is due
to its speed, flexibility and ease of computational storage requirements.
The limitation of the FDTD technique is related to the fact that the
computational domain is finite. As the result, an injected pulse experiences
spurious reflections from the domain boundaries, which leads to mixing
between the incoming and reflected waves. In order to overcome this
limitation the so-called perfectly matched layer condition has been
introduced \cite{Berenger}. However, even with this technique, a sizable
part of the incoming flux can still be reflected back \cite{Mekis}. In many
cases the separation of spurious pulses is essential for the interpretation
of the results, and this separation can only be achieved by increase of a
size of the computational domain \cite{Yu}. This may lead to a prohibitive
amount of computational work, because the stability of the FDTD algorithm
requires a sufficiently small time step.

The problem of the spurious reflections from the computational domain
boundaries does not arise in the methods based on the scattering matrix
technique, where the incident and outgoing fields are related with the help
of the scattering matrix \cite{Maystre,Whittaker,Li1,Li2,Q}. Other
approaches where the spurious reflections are avoided include e.g. a
multiple multipole method \cite{MMM}, and a Green's function method \cite%
{Green} based on the analytical expression for the Green's
function for an empty space. The main objective of the present
paper is to present a novel computational approach based on the
recursive Green's function technique that can account for an
infinite extension of a photonic crystal. In this technique the
Green's function of the photonic structure is calculated
\textit{recursively} by adding slice by slice on the basis of
Dyson's equation. In order to account for the infinite extension
of the structure both into air and into the space occupied by the
photonic crystal we make use of the so-called \textquotedblleft
surface Green's functions" that propagate the electromagnetic
fields into infinity. In this paper we present a method for
calculation of the surface Green's functions both for the case of
a semi-infinite homogeneous dielectrics, as well as for the case
of a semi-infinite periodic structure (photonic crystal). This
makes it possible to apply the Green's function technique for
investigation of a variety of important structures including
waveguides and cavities in infinite or semi-infinite photonic
crystals, as well as to study the effect of the surface states and
the coupling of waveguide Bloch modes to the external radiation.
Note that the recursive Green's function technique is widely used
for quantum mechanical transport calculations
\cite{Datta,Ferry,Sols,Z} and is proven to be unconditionally
numerically stable for various discretization schemes.

The article is organized as follows. In Section II we present a
general formulation of the problem. A description of the recursive
Green's function technique is given in Section III. This section
also provides a recipe for the calculation of Bloch states in a
periodic structure as well as the surface Green's function.
Technical details of the calculations are given in Appendices A-C.
Several examples of the application of the developed method are
given in Section IV. The conclusions are presented in Section V.

\section{General formulation of the problem}

We start with Maxwell's equations in two dimensions%
\begin{align}
\frac{1}{\varepsilon_{r}(\mathbf{r})}\mathbf{\nabla}\times \left\{ \mathbf{%
\nabla \times E(r}) \right\} & =\frac{\omega^{2}}{c^{2}}\mathbf{E}(\mathbf{r}%
)  \label{maxwell_1} \\
\mathbf{\nabla}\times\left\{ \frac{1}{\varepsilon_{r}(\mathbf{r})}\mathbf{%
\nabla \times H}(\mathbf{r})\right\} & =\frac{\omega^{2}}{c^{2}}\mathbf{H}(%
\mathbf{r}),  \notag
\end{align}
where $\mathbf{r}=x\mathbf{i}+y\mathbf{j}$, $\mathbf{\nabla }=\frac{\partial%
}{\partial x}\mathbf{i}+\frac{\partial}{\partial y}\mathbf{j}$, $%
\varepsilon_{r}(\mathbf{r})$ is the relative dielectric constant, and the
electric and magnetic field vectors $\mathbf{E(r},t)=\mathbf{E(r}%
)\exp(-i\omega t),\ \mathbf{H(r},t)=\mathbf{H(r})\exp(-i\omega t)$. If the
dielectric constant $\varepsilon_{r}(\mathbf{r})$\ is independent on $z$,
the Maxwell's equations decouple in two sets of equations for the TE modes ($%
H_{z},E_{x},E_{y}$),%
\begin{align}
& \frac{\partial}{\partial x}\frac{1}{\varepsilon_{r}}\frac{\partial }{%
\partial x}H_{z}+\frac{\partial}{\partial y}\frac{1}{\varepsilon_{r}}\frac{%
\partial}{\partial y}H_{z}+\frac{\omega^{2}}{c^{2}}H_{z}=0,
\label{TE_modes_1} \\
& E_{x}=\frac{i}{\omega\varepsilon_{0}\varepsilon_{r}}\frac{\partial H_{z}}{%
\partial y},  \notag \\
& E_{y}=\frac{-i}{\omega\varepsilon_{0}\varepsilon_{r}}\frac{\partial H_{z}}{%
\partial x},  \notag
\end{align}
\newline
and for the TM modes ($E_{z},H_{x},H_{y}$),%
\begin{align}
& \frac{1}{\varepsilon_{r}}\left( \frac{\partial^{2}E_{z}}{\partial x^{2}}+%
\frac{\partial^{2}E_{z}}{\partial y^{2}}\right) +\frac{\omega^{2}}{c^{2}}%
E_{z}=0,  \label{TM_modes_1} \\
& H_{x}=\frac{-i}{\omega\mu_{0}}\frac{\partial E_{z}}{\partial y},  \notag \\
& H_{y}=\frac{i}{\omega\mu_{0}}\frac{\partial E_{z}}{\partial x}.  \notag
\end{align}

Let us rewrite the equations for $H_{z},E_{z}$ \ (\ref{TE_modes_1}), (\ref%
{TM_modes_1})\ in an operator form \cite{Sakoda}%
\begin{equation}
\mathbf{L}f=\left( \frac{\omega }{c}\right) ^{2}f  \label{L}
\end{equation}%
where the \emph{Hermitian }differential\emph{\ }operator $\mathbf{L}$ and
the function $f$ reads,%
\begin{align}
\text{TE modes}& \text{:}\text{ }f\equiv H_{z},\ \mathbf{L}_{TE}=-\frac{%
\partial }{\partial x}\frac{1}{\varepsilon _{r}}\frac{\partial }{\partial x}-%
\frac{\partial }{\partial y}\frac{1}{\varepsilon _{r}}\frac{\partial }{%
\partial y},  \label{L_TE} \\
\text{TM modes}& \text{:}\text{ }f=\sqrt{\varepsilon _{r}}E_{z},\;\mathbf{L}%
_{TM}=-\frac{1}{\sqrt{\varepsilon _{r}}}\left( \frac{\partial ^{2}}{\partial
x^{2}}+\frac{\partial ^{2}}{\partial y^{2}}\right) \frac{1}{\sqrt{%
\varepsilon _{r}}}  \label{L_TM}
\end{align}%
\newline
For the numerical solution, Eqs. (\ref{L})-(\ref{L_TM}) have to be
discretized, $x,y\rightarrow m\Delta ,n\Delta ,$ where $\Delta $ is the grid
step. Using the following discretization of the differential operators in
Eqs. (\ref{L_TE}),(\ref{L_TM}) \cite{num_rec},%
\begin{align}
\Delta^{2}\frac{\partial }{\partial x}\xi (x)\frac{\partial f(x)}{\partial x}%
& \rightarrow \xi _{m+%
{\frac12}%
}\left( f_{m+1}-f_{m}\right) -\xi _{m-%
{\frac12}%
}\left( f_{m}-f_{m-1}\right)  \notag \\
\Delta^{2}\frac{\partial ^{2}}{\partial x^{2}}\xi (x)f(x)& \rightarrow \xi
_{m+1}f_{m+1}-2\xi _{m}f_{m}+\xi _{m-1}f_{m-1}  \label{discretization}
\end{align}%
we arrive to the finite difference equation%
\begin{align}
& v_{m,n}f_{m,n}-u_{m,m+1;n,n}f_{m+1,n}-u_{m,m-1;n,n}f_{m-1,n}-
\label{finite difference} \\
-& u_{m,m;n,n+1}f_{m,n+1}-u_{m,m;n,n-1}f_{m,n-1}=\left( \frac{\omega \Delta}{%
c}\right) ^{2}f_{m,n},  \notag
\end{align}%
where the coefficients $v,u$ are defined for the cases of TE and TM modes as
follows
\begin{align}
\text{TE modes}& \text{: }f_{m,n}=H_{z\,m,n};\;\xi _{m,n}=\frac{1}{%
\varepsilon _{r\,m,n}},  \label{TE_coefficients} \\
v_{m,n}& =\xi _{m+{\frac{1}{2}},n}+\xi _{m-{\frac{1}{2}},n}+\xi _{m,n+{\frac{%
1}{2}}}+\xi _{m,n-{\frac{1}{2}}},  \notag \\
u_{m,m+1;n,n}& =\xi _{m+{\frac{1}{2}},n},\;u_{m,m-1;n,n}=\xi _{m-{\frac{1}{2}%
},n},  \notag \\
u_{m,m;n,n+1}& =\xi _{m,n+{\frac{1}{2}}},\;u_{m,m;n,n-1}=\xi _{m,n-{\frac{1}{%
2}}};  \notag
\end{align}

\begin{align}  \label{TM_coefficients}
\text{TM modes}& \text{:}f_{m,n}=\sqrt{\varepsilon _{r\,m,n}}%
E_{z\,m,n};\;\xi _{m,n}=\frac{1}{\sqrt{\varepsilon _{r\,m,n}}} \\
v_{m,n}& =4\xi _{m,n}^{2},  \notag \\
u_{m,m+1;nn}& =\xi _{m,n}\xi _{m+1,n},\;u_{m,m-1;nn}=\xi _{m-1,n}\xi _{m,n},
\notag \\
u_{m,m;n,n+1}& =\xi _{m,n+1}\xi _{m,n},\;u_{m,m;n,n-1}=\xi _{m,n}\xi_{m,n-1}.
\notag
\end{align}

A convenient and common way to describe finite-difference equations on a
numerical grid (lattice) is to introduce the corresponding \textit{%
tight-binding }operator\textit{.} For this purpose we first introduce
creation and annihilation operators, $a_{m,n}^{+}$, $a_{m,n}.$ Let the state
$|0\rangle \equiv |0,\ldots ,0_{m,n},\ldots ,0\rangle $\ describe an empty
lattice, and the state $|0,\ldots 0,1_{m,n},0,\ldots ,0\rangle $\ describe
an excitation at the site $m,n$. The operators $a_{m,n}^{+}$, $a_{m,n}$ act
on these states according to the rules \cite{Ferry}
\begin{align}
& a_{m,n}^{+}|0\rangle =|0,\ldots 0,1_{m,n},0,\ldots ,0\rangle ,  \label{a+}
\\
& a_{m,n}^{+}|0,\ldots 0,1_{m,n},0,\ldots ,0\rangle =0,  \notag
\end{align}%
\newline
and
\begin{align}
& a_{m,n}|0\rangle =0,  \label{a-} \\
& a_{m,n}|0,\ldots 0,1_{m,n},0,\ldots ,0\rangle =|0\rangle .  \notag
\end{align}%
and they obey the following commutational relations%
\begin{align}
\lbrack a_{m,n},a_{m,n}^{+}]& =a_{m,n}a_{m,n}^{+}-a_{m,n}^{+}a_{m,n}=\delta
_{m,n};  \label{com} \\
\lbrack a_{m,n},a_{m,n}]& =[a_{m,n}^{+},a_{m,n}^{+}]=0.  \notag
\end{align}%
Consider an operator equation%
\begin{equation}
\widehat{\mathcal{L\,}}|f\rangle =\left( \frac{\omega \Delta}{c}\right)
^{2}|f\rangle,  \label{L_operator_2}
\end{equation}%
where the Hermitian operator%
\begin{align}
~\widehat{\mathcal{L}}& =\sum_{m,n}(v_{m,n}a_{m,n}^{+}a_{m,n}-
\label{L_operator} \\
& -u_{m,m+1;n,n}a_{m,n}^{+}a_{m+1,n}-u_{m+1,m;n,n}a_{m+1,n}^{+}a_{m,n}-
\notag \\
& -u_{m,m;n,n+1}a_{m,n}^{+}a_{m,n+1}-u_{m,m;n+1,n}a_{m,n+1}^{+}a_{m,n})
\notag
\end{align}%
acts on the state%
\begin{equation}
|f\rangle =\sum_{m,n}f_{m,n}a_{m,n}^{+}|0\rangle .  \label{f_state}
\end{equation}%
Substituting the above expressions for $\widehat{\mathcal{L}}$ and $%
|f\rangle $ in Eq. (\ref{L_operator_2}), and using the commutation
relations and the rules Eqs. (\ref{a+})-(\ref{com}), it is
straightforward to demonstrate that the operator equation
(\ref{L_operator_2}) is equivalent to the finite difference
equation (\ref{finite difference}). Note an apparent physical
meaning of the last four terms in Eq. (\ref{L_operator}): terms 2
and 3 describe forward and backward hopping between two
neighboring sites in the $x$-direction, and terms 4 and 5 denote
similar hopping in the $y$-direction. In the next section we
outline the Green's function formalism for solution of Eq.
(\ref{L_operator_2}).

\section{The recursive Green's function technique}

\subsection{Basics}

Let us first specify structures under investigation. We consider light
propagation through a photonic structure defined in a waveguide (supercell)
of the width $N,$ where we assume the cyclic boundary condition (i.e. the
row $n=N+1$ coincides with the row $n=1$). The photonic structure occupies a
finite internal region consisting of $M$ slices ($1\leq m\leq M$).

\begin{figure}[ptb]
\begin{center}
\includegraphics[
scale=0.7 ]{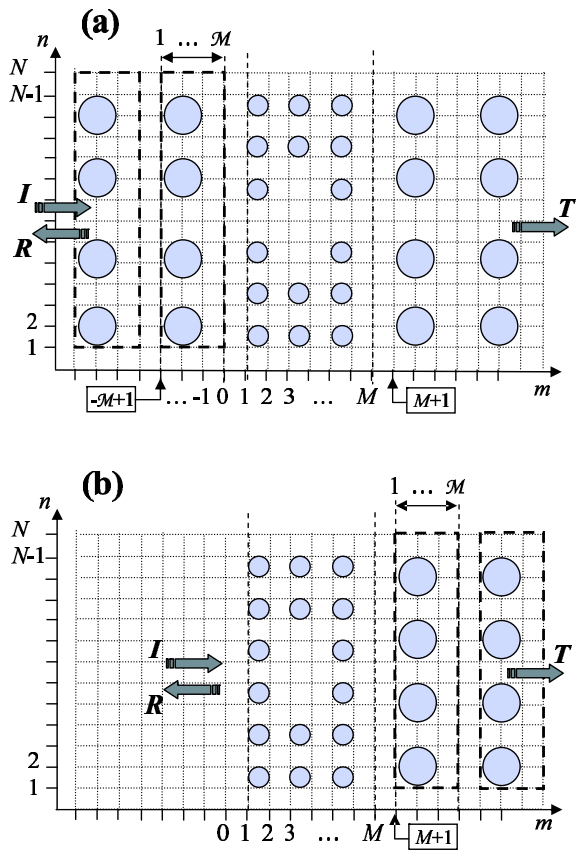}
\end{center}
\caption{Schematic illustration of the system under study defined
in a waveguide (supercell) of the width \textit{N}. An internal
region of the structure occupies \textit{M} slices. Two
representative cases are shown, (a) external regions are
semi-periodic photonic crystals with the period $\mathcal{M}$, (b)
external regions represent a semi-infinite periodic photonic
crystal with the period $\mathcal{M}$ to the right and air to the
left. Arrows indicate the directions on the incoming (\textit{I}),
reflected (\textit{R}), and transmitted (\textit{T}) waves.}
\label{fig1_schema}
\end{figure}

The external regions are semi-infinite waveguides (supercells)
extending into regions $m\leq 0$ and $m\geq M+1$. The waveguides
can represent air (or a material with a constant refractive
index), or a periodic photonic crystal. Figure \ref{fig1_schema}
shows two representative examples where (a) the semi-infinite
waveguides represent a \textit{periodic} photonic crystal with the
period $\mathcal{M}$, and (b) a photonic structure is defined at
the boundary between air and the semi-infinite photonic crystal.

Let us first define the scattering states for the structures under
consideration. The translation invariance along the supercell dictates the
Bloch form for the $\alpha $th incoming state $|\psi _{\alpha }^{\mathrm{i}%
}\rangle $,
\begin{equation}
|\psi _{\alpha }^{\mathrm{i}}\rangle =\sum_{m\leqq 0}e^{ik_{\alpha
}^{+}m}\sum_{n=1}^{N}\phi _{m,n}^{\alpha }\,a_{m,n}^{+}|0\rangle ,
\label{s_Bloch}
\end{equation}%
where $k_{\alpha }^{+}\,(k_{\alpha }^{-})$ is the Bloch wave
vector of the right-propagating (left-propagating) state $\alpha
$, and $\phi _{m,n}^{\alpha }$is the corresponding Bloch
transverse eigenfunction
satisfying the Bloch condition%
\begin{equation}
\phi _{m,n}^{\alpha }=\phi _{m+\mathcal{M},n}^{\alpha }.  \label{Bloch}
\end{equation}%
The transmitted and reflected states, $|\psi _{\alpha }^{\mathrm{t}%
}\rangle $ and $|\psi _{\alpha }^{\mathrm{r}}\rangle $, can be
written in a similar form,
\begin{align}
|\psi _{\alpha }^{\mathrm{t}}\rangle & =\sum_{m\geqq M+1}\sum_{\beta
}t_{\beta \alpha }e^{ik_{\beta }^{+}(m-(M+1))}\sum_{n=1}^{N}\phi
_{m,n}^{\beta }\,a_{m,n}^{+}|0\rangle ,  \label{t} \\
|\psi _{\alpha }^{\mathrm{r}}\rangle & =\sum_{m\leqq 0}\sum_{\beta }r_{\beta
\alpha }e^{ik_{\beta }^{-}m}\sum_{n=1}^{N}\phi _{m,n}^{\beta
}\,a_{m,n}^{+}|0\rangle ,  \label{r}
\end{align}%
where $t_{\beta \alpha }\,(r_{\beta \alpha })$ stands for the transmission
(reflection) amplitude from the incoming Bloch state $\alpha $ to the
transmitted (reflected) Bloch state $\beta .$ Note that in general case the
wave vectors $k_{\alpha }^{\pm }$ and the Bloch states $\phi _{m,n}^{\alpha
} $ can be different in the left and right waveguides (see e.g. Fig. \ref%
{fig1_schema}(a), when the photonic structure is defined at the
boundary air/photonic crystal). The method of calculation of the
Bloch states for an arbitrary periodic structure is described
below in Section IIIC.

\bigskip\ We define Green's function of the operator $\widehat{\mathcal{L\,}}
$ in a standard way,
\begin{equation}
\left( \left( \omega \Delta /c\right) ^{2}-\widehat{\mathcal{L\,}}\right)
G(\omega )=\widehat{1\mathcal{\,}},  \label{Greens function}
\end{equation}%
where $\widehat{1\mathcal{\,}}$ is the unitary
operator\cite{Economou}. The knowledge of the Green's function
allows one to calculate the transmission and reflection
coefficients. Indeed, let us write down the solution of Eq. (\ref%
{L_operator_2}) as a sum of two terms, the incoming state $|\psi ^{\mathrm{i}%
}\rangle $ and the system response $|\psi \rangle $ representing whether the
transmitted or reflected states, $|\psi ^{\mathrm{t}}\rangle $ or $|\psi ^{%
\mathrm{r}}\rangle $, $|f\rangle =|\psi ^{\mathrm{i}}\rangle +|\psi \rangle
. $ Substituting $|f\rangle $ into Eq. (\ref{L_operator_2}) and using the
formal definition of the Green's function Eq. (\ref{Greens function}), the
solution of Eq. (\ref{L_operator_2}) can be written in the form

\begin{equation}  \label{respons}
|\psi \rangle =G\left( \widehat{\mathcal{L\,}}-\left( \omega \Delta
/c\right) ^{2}\right) |\psi ^{\mathrm{i}}\rangle.
\end{equation}

Calculating the matrix elements $\langle M+1,n|\psi \rangle \equiv \langle
0|a_{M+1,n}\psi \rangle $ and $\langle 0,n|\psi \rangle \equiv \langle
0|a_{0,n}\psi \rangle ,$ of the rigth and left hand side of Eq. (\ref%
{respons}), we arrive to the $N\times N$ system of linear equations for the
transmission and reflection amplitudes (see for details Appendix A),
\begin{align}
\Phi _{M+1}T& =-G^{M+1,0}(U_{0,1}\Phi _{-\mathcal{M+}1}K_{l}-{\Gamma _{l}}%
^{-1}\Phi _{0})  \label{T} \\
\Phi _{0}R& =-G^{0,0}(U_{0,1}\Phi _{-\mathcal{M+}1}K_{l}-{\Gamma _{l}}%
^{-1}\Phi _{0})-\Phi _{0}  \label{R}
\end{align}%
where the matrix elements $(T)_{\beta \alpha }=t_{\beta \alpha },$ $%
(R)_{\beta \alpha }=r_{\beta \alpha };$ $G^{M+1,0}$ and $G^{0,0}$ are the
Green's function matrixes with the elements
\begin{equation}
(G^{m,l})_{n,p}=\langle 0|a_{m,n}G\,a_{l,p}^{+}|0\rangle .
\label{matrix_element}
\end{equation}%
$\Gamma _{l}\equiv G_{\mathrm{wg}}^{0,0}$ is the left \textquotedblleft
surface Green's function\textquotedblright\ corresponding only to part of
the whole structure, namely, to the semi-infinite waveguide (supercell) that
extends to the left, $-\infty <m\leq 0.$ The physical meaning of the surface
Green's function $\Gamma $ is that it propagates the electromagnetic fields
from the boundary slice of the semi-infinite waveguide (supercell) into
infinity. A method for calculation of the surface Green's functions both for
the case of a semi-infinite homogeneous dielectrics, as well as for the case
of a semi-infinite photonic crystal in a waveguide geometry is described
below in Section IIID. The matrixes $K_{l}$ and $\Phi _{m}$ are given by the
right-propagating Bloch eigenvectors $k_{\alpha }^{+}\,$and the
corresponding eigenstates $\phi _{m,n}^{\alpha }$ in the waveguides,
\begin{equation}
(K_{l})_{\alpha \beta }=\exp (ik_{\alpha }^{+})\,\delta _{\alpha \beta
};(\Phi _{m})_{n\alpha }=\phi _{m,n}^{\alpha }\;  \label{K_matrix}
\end{equation}%
and the diagonal \textquotedblleft hopping matrix\textquotedblright\ $%
U_{0,1} $ is defined as

\begin{equation}
(U_{0,1})_{n,n^{\prime }}=u_{0,1;n,n^{\prime }}\delta _{n,n^{\prime }}.
\label{U}
\end{equation}%
(Note that the matrix $K_{l}$ in Eqs. (\ref{T}),(\ref{R}) refers to the
right-propagating states in the \textit{left }waveguide). In the following
sections we describe the recursive Green's function technique based on the
successive use of the Dyson's equation, introduce the method for the
calculation of Bloch states in a periodic structure, and outline the way to
calculate the surface Green's function $\Gamma .$

\subsection{Recursive technique based on Dyson's equations}

\begin{figure}[ptb]
\begin{center}
\includegraphics[scale = 0.6]{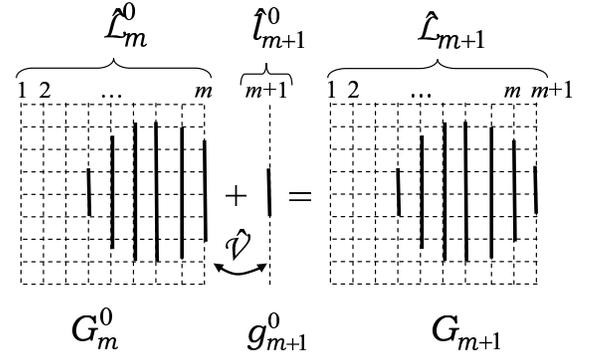}
\end{center}
\caption{Schematic illustration of the application of Dyson's equation for
calculation of Green's function for a composed structure consisting of $m$+1
slices (see text for details). The operators $\widehat{\mathcal{L\,}}%
_{m}^{0} $ and $\widehat{\mathit{l}\mathcal{\,}}_{m+1}^{0}$ describe
respectively the structure composed of $m$ slices, and the ($m+1$)-th slice.
The operator $\widehat{\mathcal{L\,}}_{m+1}=\widehat{\mathcal{L\,}}_{m}^{0}+%
\widehat {\mathit{l}\mathcal{\,}}_{m+1}^{0}+\widehat{\mathcal{V\,}}$
corresponds to the composed structure of $m+1$ slices, where $\widehat{%
\mathcal{V\,}}$ is the perturbation operator describing the hopping between the $%
m$th and ($m+1$)-th slices.}
\label{fig2_Dyson}
\end{figure}

In order to calculate Green's function of the internal structure (i.e. for
the slices $1\leq m\leq M$) we utilize the recursive technique based on
Dyson's equation, see Fig. \ref{fig2_Dyson}.

In order to illustrate this technique let us consider a structure consisting
of $m$ slices\bigskip . The operator $\widehat{\mathcal{L\,}}_{m}^{0}$
describing this structure can be written down in the form%
\begin{equation}
\widehat{\mathcal{L\,}}_{m}^{0}=\sum_{\mathbf{r}}v_{\mathbf{r}}a_{\mathbf{r}%
}^{+}a_{\mathbf{r}}-\sum_{\mathbf{r,\Delta }}u_{\mathbf{r,r+\Delta }}a_{%
\mathbf{r}}^{+}a_{\mathbf{r+\Delta }},  \label{L_m}
\end{equation}%
where $\mathbf{r}=m^{\prime },n^{\prime }\ (1\leq m^{\prime }\leq m;1\leq
n^{\prime }\leq N),$ and the summation over $\Delta $ in the second term is
performed over all available nearest neighbors. Suppose we know Green's
function $G_{m}^{0}$ of the operator $\widehat{\mathcal{L\,}}_{m}^{0}$, as
well as Green's function $g_{m+1}^{0}$ of the the operator $\widehat{\mathit{l%
}\mathcal{\,}}_{m+1}^{0}$ corresponding to a single ($m+1$)-th
slice,
\begin{align}
~\widehat{\mathit{l}\mathcal{\,}}_{m+1}^{0}&
=\sum_{n}(v_{m+1,m+1}a_{m+1,n}^{+}a_{m+1,n}-  \label{l_m+1} \\
& -u_{m+1,m+1;n,n+1}a_{m+1,n}^{+}a_{m+1,n+1}-  \notag \\
& -u_{m+1,m+1;n+1,n}a_{m+1,n+1}^{+}a_{m+1,n}).  \notag
\end{align}%
(The method of calculation of Green's function for a single slice
is outlined in Appendix C). Our aim is to calculate Green's
function of the composed structure, $G_{m+1}$, consisting of $m+1$
slices. The operator
corresponding to this structure can be written down in the form%
\begin{equation}
\widehat{\mathcal{L\,}}_{m+1}=\widehat{\mathcal{L\,}}_{m}^{0}+\widehat{l%
\mathcal{\,}}_{m+1}^{0}+\widehat{\mathcal{V\,}},  \label{L_m+1}
\end{equation}%
where the operators $\widehat{\mathcal{L\,}}_{m}^{0}$ and $\widehat{l%
\mathcal{\,}}_{m+1}^{0}$ are given by the expressions Eqs. (\ref{L_m+1}), (%
\ref{l_m+1}), and $\widehat{\mathcal{V\,}}=$ $\widehat{\mathcal{V\,}}%
_{m,m+1}+$ $\widehat{\mathcal{V\,}}_{m+1,m}$ is the perturbation
operator describing the hopping between the $m$th and ($m+1$)-th
slices,
\begin{align}
~\widehat{\mathcal{V\,}}& =\widehat{\mathcal{V\,}}_{m+1,m}+\widehat{\mathcal{%
V\,}}_{m,m+1}=  \label{V} \\
& =-\sum_{n}(u_{m,m+1;n,n}a_{m,n}^{+}a_{m+1,n}+  \notag \\
& +u_{m+1,m;n,n}a_{m+1,n}^{+}a_{m,n}).  \notag
\end{align}%
The Green's function of the composed structure, $G_{m+1},$ can be calculated
on the basis of Dyson's equation\cite{Economou}%
\begin{align}
G_{m+1}& =G^{0}+G^{0}\widehat{\mathcal{V}}G_{m+1},  \label{dyson} \\
G_{m+1}& =G^{0}+G_{m+1}\widehat{\mathcal{V}}G^{0},  \notag
\end{align}%
where $G^{0}$ is the `unperturbed' Green's function corresponding to the
operators $\widehat{\mathcal{L\,}}_{m}^{0}$ or $\widehat{l\mathcal{\,}}%
_{m+1}^{0}$. For the sake of completeness, a brief derivation of Dyson's
equation is given in Appendix B. Thus, starting from Green's function for
the first slice $g_{1}^{0}$ and adding recursively slice by slice we are in
the position to calculate Green's function of the internal structure
consisting of $M$ slices. Explicit expressions following from Eqs. (\ref%
{dyson}) and used for the recursive calculations are given below,
\begin{align}
G_{m+1}^{m+1,m+1}&
=(I-g_{m+1}^{0}U_{m+1,m}(G_{m}^{0})^{m,m}U_{m,m+1})^{-1}g_{m+1}^{0},  \notag
\label{dyson_explicit} \\
& \\
G_{m+1}^{m+1,1}& =G_{m+1}^{m+1,m+1}U_{m+1,m}(G_{m}^{0})^{m,1},  \notag \\
&  \notag \\
G_{m+1}^{1,1}& =(G_{m}^{0})^{1,1}+(G_{m}^{0})^{1,m}U_{m,m+1}G_{m+1}^{m+1,1},
\notag \\
&  \notag \\
G_{m+1}^{1,m+1}& =(G_{m}^{0})^{1,m}U_{m,m+1}G_{m+1}^{m+1,m+1},  \notag
\end{align}%
where the upper indexes define the matrix elements of the Green's function $%
G^{m,m^{\prime }}=\langle 0|a_{m,n}G\,a_{m^{\prime },n^{\prime
}}^{+}|0\rangle $. This recursive technique is proven to be unconditially
numerically stable \cite{Datta,Ferry,Sols}. The performance of the method is
determined by the size of the system of linear equations (\ref%
{dyson_explicit}) which we solve when we add each consecutive
slice. This system is solved $M$ times, where $M$ is the number of
slices of the internal structure (in the $x$-direction).  The size
of Eqs. (\ref{dyson_explicit}) is $N\times N$, where $N$ is a
number of discretization points in the $y$-direction. Typical
dimensions of the equations used for computations of the
structures reported in Section 4 are $\thicksim 200\times 200.$

In order to calculate the Green's function of the whole system, we have to
connect the internal structure with the left and right semi-infinite
waveguides. Starting with the left waveguide, we write%
\begin{equation}
\widehat{\mathcal{L\,}}_{int+left}=\widehat{\mathcal{L\,}}_{int}+\widehat{%
\mathcal{L\,}}_{left}+\widehat{\mathcal{V\,}},  \label{left_wg}
\end{equation}%
where the operators $\widehat{\mathcal{L\,}}_{int+left},\widehat{\mathcal{L\,%
}}_{int}$ and $\widehat{\mathcal{L\,}}_{left}$ describe respectively the
system representing the internal structure + the left waveguide, the
internal structure, and the left waveguide. The perturbation operator $%
\widehat{\mathcal{V}}\mathcal{\,}$ describes the hopping between the left
waveguide and the internal structure. Applying then the Dyson equation in a
similar way as we described above,
\begin{equation}
G_{int+left}=G^{0}+G^{0}\widehat{\mathcal{V}}G_{int+left},
\label{Green_left}
\end{equation}%
we are in the position to find the Green's function $G_{int+left}$ of the
system representing the internal structure + the left waveguide. $G^{0}$ in
Eq. (\ref{Green_left}) in an `unperturbed' Green's function corresponding to
the internal structure and the semi-infinite waveguide (the
\textquotedblleft surface Green's function\textquotedblright\ $\Gamma ).$
Having calculated the Green's function $G_{int+left}$ on the basis of Eq. (%
\ref{Green_left}), we proceed in a similar way by adding the right waveguide
and calculating with the help of the Dyson's equation the total Green's
function $G$ of the whole system.

\subsection{Bloch states of the periodic structure}

In this section we describe the method for calculation of the
Bloch states in periodic waveguides (supercells) using the Green's
function technique. Similar method was used for calculation of
Bloch states in quantum-mechanical structures\cite{Z}.

\begin{figure}[ptb]
\begin{center}
\includegraphics[
scale=0.7 ]{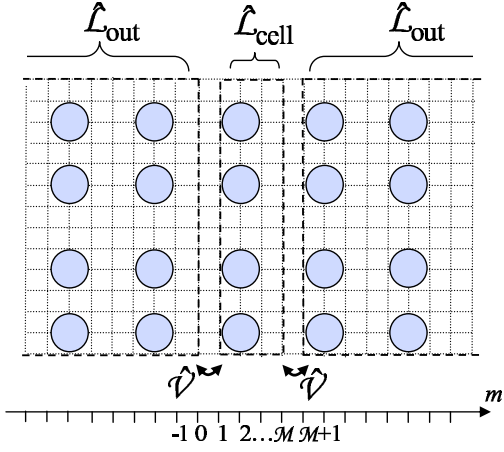}
\end{center}
\caption{Schematic illustration of the calculation of Bloch states in an
infinite periodic structure (see text for details). The operator $\widehat{%
\mathcal{L\,}}_{\mathrm{cell}}$ describes a unit cell under consideration, $%
1\leq m\leq\mathcal{M}$, and $\widehat{\mathcal{L\,}}_{\mathrm{out}}$
describes the rest of the structure. The hopping between the cell and the
rest of the structure is described by the operator $\widehat {\mathcal{V\,}}$%
.}
\label{fig3_Bloch}
\end{figure}

Consider a unit cell of a periodic waveguide occupying $\mathcal{M}$ slices,
$1\leq m\leq\mathcal{M}$, see Fig. \ref{fig3_Bloch}.

Rewrite the operator corresponding to the whole structure in the form%
\begin{equation}
\widehat{\mathcal{L\,}}=\widehat{\mathcal{L\,}}_{\mathrm{cell}}+\widehat{%
\mathcal{L\,}}_{\mathrm{out}}+\widehat{\mathcal{V\,}},  \label{L_cell}
\end{equation}%
where the operators $\widehat{\mathcal{L\,}}_{\mathrm{cell}}$ and $\widehat{%
\mathcal{L\,}}_{\mathrm{out}}$ describe respectively the cell under
consideration ($1\leq m\leq \mathcal{M}$), and the outside region including
all other slices $-\infty <m\leq 0$ and $\mathcal{M+}1\leq m<\infty $, and $%
\widehat{\mathcal{V\,}}$ is the hopping operator between the cell and slices
$m=0$ and $m=\mathcal{M}+1$. Write the total wave function $|\psi \rangle
=\sum_{m,n}\psi _{m,n}a_{m,n}^{+}|0\rangle $ in the form%
\begin{equation}
|\psi \rangle =|\psi _{\mathrm{cell}}\rangle +|\psi _{\mathrm{out}}\rangle ,
\label{psi_cell}
\end{equation}%
where $|\psi _{\mathrm{cell}}\rangle $ and $|\psi _{\mathrm{out}}\rangle $
are respectively wave functions in the cell and in the outside region.
Substituting Eqs. (\ref{L_cell}),(\ref{psi_cell}) into Eq. (\ref%
{L_operator_2}), we obtain $|\psi _{\mathrm{cell}}\rangle =G_{\mathrm{cell}}%
\widehat{\mathcal{V\,}}|\psi _{\mathrm{out}}\rangle ,$ where $G_{\mathrm{cell%
}}$ is the Green's function of the operator $\widehat{\mathcal{L\,}}_{%
\mathrm{cell}}$. Calculating the matrix elements $\langle 1,n|\psi \rangle $
and $\langle \mathcal{M},n|\psi \rangle $, this equation can be written in
the matrix form,
\begin{subequations}
\begin{align}
\psi _{1}& =G_{\mathrm{cell}}^{1,1}U_{1,0}\psi _{0}+G_{\mathrm{cell}}^{1,%
\mathcal{M}}U_{1,0}\psi _{\mathcal{M}+1}  \label{psi_cell_2} \\
\psi _{\mathcal{M}}& =G_{\mathrm{cell}}^{\mathcal{M},1}U_{1,0}\psi _{0}+G_{%
\mathrm{cell}}^{\mathcal{M},\mathcal{M}}U_{1,0}\psi _{\mathcal{M}+1},
\end{align}%
where the vector column $\psi _{m}=$ $\left( \psi _{m,1}\ldots \psi
_{m,N}\right) ^{T}$, and where we used $U_{\mathcal{M},\mathcal{M}%
+1}=U_{0,1} $ (because of the periodicity) and $U_{0,1}=U_{1,0}$ (according
to the definition of $U$, Eq.(\ref{U})). It is convenient to rewrite Eq. (%
\ref{psi_cell_2}) in a compact form
\end{subequations}
\begin{align}
& T_{1}\left(
\begin{array}{c}
\psi _{\mathcal{M}+1} \\
\psi _{\mathcal{M}}%
\end{array}%
\right) =T_{2}\left(
\begin{array}{c}
\psi _{1} \\
\psi _{0}%
\end{array}%
\right) ,\;\mathrm{where}  \label{T1_T2} \\
& T_{1}=%
\begin{pmatrix}
-G_{\mathrm{cell}}^{1,\mathcal{M}}U_{1,0} & \;0 \\
G_{\mathrm{cell}}^{\mathcal{M},\mathcal{M}}U_{1,0} & \;I%
\end{pmatrix}%
,\;T_{2}=%
\begin{pmatrix}
-I & \;-G_{\mathrm{cell}}^{1,1}U_{1,0} \\
0 & \;G_{\mathrm{cell}}^{\mathcal{M},1}U_{1,0}%
\end{pmatrix}%
,  \notag
\end{align}%
\newline
with $I$ being the unitary matrix. The wave function of the periodic
structure has Bloch form,
\begin{equation}
\psi _{\mathcal{M}+m}=e^{ik_{x}\mathcal{M}}I\psi _{m}.  \label{Bloch_2}
\end{equation}%
Combining Eqs. (\ref{T1_T2}) and (\ref{Bloch_2}), we arrive to the
eigenequation for Bloch wave vectors and Bloch states,%
\begin{equation}
T_{1}^{-1}T_{2}\left(
\begin{array}{c}
\psi _{1} \\
\psi _{0}%
\end{array}%
\right) =e^{ik_{x}\mathcal{M}}\left(
\begin{array}{c}
\psi _{1} \\
\psi _{0}%
\end{array}%
\right) ,  \label{Bloch_eigenequation}
\end{equation}%
determining the set of Bloch eigenvectors $k_{x}^{\alpha }$ and
eigenfunctions $\psi ^{\alpha },$ $1\leq \alpha \leq N.$

To improve numerical stability of Eq. (\ref{Bloch_eigenequation}), it may be
rewritten in the form \cite{Li2}:
\begin{equation}
(T_{1}+T_{2})^{-1}T_{1}\left(
\begin{array}{c}
\psi _{1} \\
\psi _{0}%
\end{array}%
\right) =(e^{ik_{x}\mathcal{M}}+1)^{-1}\left(
\begin{array}{c}
\psi _{1} \\
\psi _{0}%
\end{array}%
\right).  \label{Bloch_eigenequation_modified}
\end{equation}

This technique allows one to avoid overflows and underflows in the
eigensolver routine when eigenvalues with $|e^{ik_{x}\mathcal{M}}| \gg 1$
and $|e^{ik_{x}\mathcal{M}}| \ll 1$ are calculated.

In order to separate the left- and right-propagating states we compute the
Poynting vector integrated over transverse direction, whose sign determines
the direction of propagation. Bloch state propagating in a waveguide
(supercell) defined in a photonic crystal is illustrated below in Fig. \ref%
{fig5_band_str}(c).

Poynting vector can be expressed as follows \cite{Sakoda}
\begin{equation}
\mathbf{S}_{\alpha }(y)=\frac{1}{2}\Re \lbrack \mathbf{E}_{\alpha }(y)\times
\mathbf{H}_{\alpha }^{\ast }(y)].  \label{poynt_vect}
\end{equation}
Note that for the case of the waveguide defined in air,
$\mathcal{M}=1$, and Green's functions $G_{\mathrm{cell}}$ in Eq.
(\ref{T1_T2}) is simply given by Green's function of a single
slice $g^{0}$ (see Appendix C for details of calculation of
$g^{0}$).

\subsection{The surface Green's function $\Gamma.$}

Consider a semi-infinite Bloch waveguide (supercell) of the
periodicity $\mathcal{M}$
extending in the region $-M\leq m < \infty$ as depicted in Fig. \ref%
{fig4_Gamma}

\begin{figure}[tbp]
\begin{center}
\includegraphics[
scale = 0.7 ]{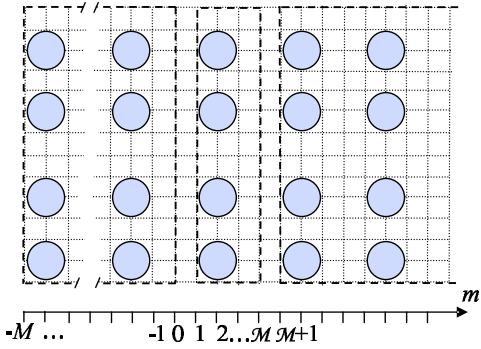}
\end{center}
\caption{A schematic diagram illustrating calculation of the surface Green's
function $\Gamma $ of a periodic structure (see text for details).}
\label{fig4_Gamma}
\end{figure}
Suppose that an excitation $|s\rangle $ is applied to its first slice $m=-M$%
. Introducing the Green function $G_{\mathrm{wg}}$ corresponding to the
operator $\widehat{\mathcal{L\,}}_{\mathrm{wg}}$ describing the waveguide,
one can write down the response to the excitation $|s\rangle $ in the form
\begin{equation}
|\psi \rangle =G_{\mathrm{wg}}|s\rangle ,  \label{psi}
\end{equation}%
where $|\psi \rangle $ is the wave function that has to satisfy Bloch
conditions (\ref{Bloch_2}). Applying Dyson's equation between the slices $0$
and $1$ we obtain%
\begin{equation}
G_{\mathrm{wg}}^{1,-M}=\Gamma _{r}U_{1,0}G_{\mathrm{wg}}^{0,-M},
\label{G_psi}
\end{equation}%
where $\Gamma _{r}\equiv G_{\mathrm{wg}}^{1,1}$ is the right surface Green's
function. (Note that because the waveguide is infinitely long and periodic, $%
G_{\mathrm{wg}}^{1,1}=G_{\mathrm{wg}}^{\mathcal{M}+1,\mathcal{M}+1}=G_{%
\mathrm{wg}}^{2\mathcal{M}+1,2\mathcal{M}+1}=$ ... etc.). Taking the matrix
elements $\langle 1,n|\psi \rangle $ of Eq. (\ref{psi}) and making use of
Eq. (\ref{G_psi}), we obtain for an each Bloch state $\alpha $, $\psi
_{1}^{\alpha }=$ $\Gamma _{r}U_{1,0}\psi _{0}^{\alpha }.$ The latter
equation can be used for determination of $\Gamma _{r}$,%
\begin{equation}
\Gamma _{r}U_{1,0}=\Psi _{1}\Psi _{0}^{-1},  \label{Gamma}
\end{equation}%
where $\Psi _{1}$ and $\Psi _{0}$ are the square matrixes composed of
matrix-columns $\psi _{1}^{\alpha }$ and $\psi _{0}^{\alpha },$ Eq. (\ref%
{Bloch_eigenequation}). If the waveguide is open to the left, its
surface Green's function is the same as the surface Green's
function of the corresponding waveguide open to the right, $\Gamma
_{l}=\Gamma _{r}.$ Note that for the case of the waveguide defined
in air the surface Green's function (\ref{Gamma}) simplifies to
$\Gamma _{r}U_{1,0}=K$, where $K$ is defined according to Eq.
(\ref{K_matrix}).

\section{Applications of the method}

To reveal the power of the method we study three model systems defined in 2D
square-lattice photonic crystal. First, we calculate a transmission
coefficient and quality factor ($Q$ factor) of several representative types
of microcavities in infinite PCs. Then we focus on semi-infinite crystals
where we investigate the effect of surface states, and, finally, we consider
a semi-infinite PC with a waveguide opening to the surface. For the bulk
crystal we choose a structure composed of cylindrical rods with the
permittivity $\varepsilon _{r}=8.9$ and the diameter of a rod $d=0.4a$ in a
vacuum background, where $a$ is the size of the unit cell. Each unit cell is
discretized into 25 points in both $x$ and $y$ directions.

Most of photonic crystal devices operate in a bandgap. The structure at hand
has a complete bandgap for TM-modes in the frequency range $0.32\lesssim
\omega a / 2\pi c \lesssim 0.44$ \cite{JP}, and does not have a complete
bandgap for the TE-polarization. Because of this, we will hereafter consider
the TM-modes only.

\begin{figure}[tbp]
\begin{center}
\includegraphics[scale = 0.5]{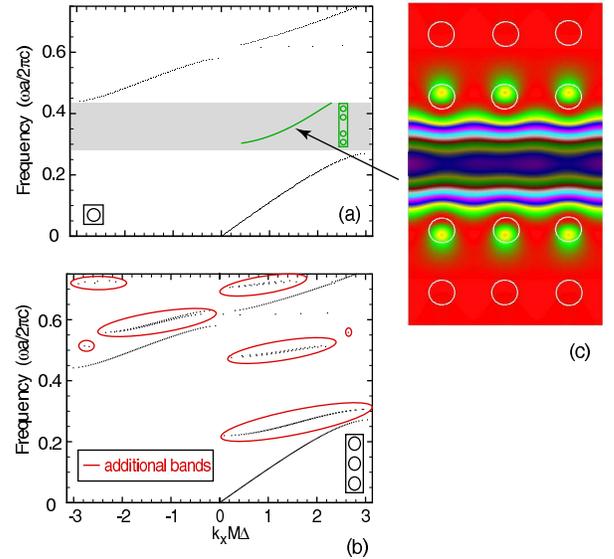}
\end{center}
\caption{(color online) (a) Band diagram for the right-propagating TM-mode
of an infinite 2D photonic crystal ($\protect\varepsilon _{r}=8.9$, $d=0.4a$%
) in $\Gamma X$-direction. PC has a fundamental bandgap in the
frequency range $0.28\lesssim \protect\omega a / 2\protect\pi C
\lesssim 0.44$ (filled with gray in the figure). Green line in the
fundamental bandgap corresponds a guided mode in a waveguide
created by removing a central row of rods from the PC as shown in
the inset. (b) Additional bands (encircled with red) originated
from the finite size effect. The waveguide (supercell) contains
three unit cells in the transverse direction as illustrated in the
inset.
(c) Bloch state propagating in the PC waveguide at $\protect%
\omega a / 2\protect\pi c =0.38$}
\label{fig5_band_str}
\end{figure}

The developed method allows one to treat structures unlimited in $x$%
-direction, whereas in $y$-direction the structure of interest is
confined within a supercell with imposed cyclic boundary
conditions. This leads to the finite size effects in a photonic
band structure. If the supercell consists more than one elementary
cell, additional bands appear along with the bands for infinite PC
(Fig. \ref{fig5_band_str} (a,b)), as the result of the imposed
boundary conditions in the transverse direction.

\begin{figure}
\begin{center}
\includegraphics[
scale = 0.35
]{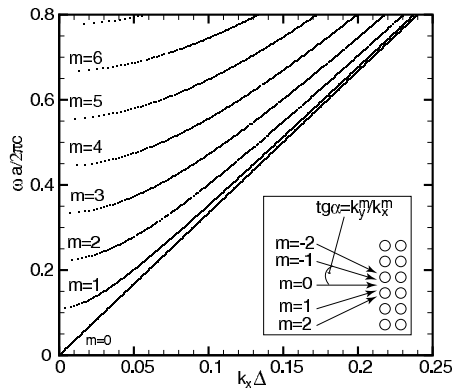}
\end{center}
\caption{Dispersion relation for the air supercell of the width of
9$a$. Effective angles of incidence are determined by the angular
wavenumber $m$. Inset shows the effective angles of incidence for
$m=-2,-1,0,1,2$.} \label{fig6_band_str_air}
\end{figure}

A similar finite size effect emerges when air waveguides
(supercells) are attached to the system of interest. Even though
we send a wave from an open space, we use a finite number of
propagating modes. Solution of the eigenvalue problem (\ref{L})
for the air supercell gives a discrete set of right-propagating
eigenstates $k_{x}^{m}=\sqrt{\omega ^{2}/c^{2}-(2\pi m/w)^{2}}$
where $w$ is the width of the supercell, and $m$ is integer such
that $\max{|m|}<\omega w/2\pi c$. Thus, a wave incident from air
effectively propagates only at certain incidence angles,
determined by the ratio of the longitudinal and transverse wave
vectors $\tan \alpha =k_{y}^{m}/k_{x}^{m}$, as illustrated in Fig.
\ref{fig6_band_str_air}. Note that this finite size effect (caused
by the cyclic boundary conditions in the $y$-direction) might in
some cases represent a drawback of the method.

\subsection{Microcavity}

In this section we consider a microcavity defined in a waveguide
in an infinite PC. The waveguide is created by removing a single
central row of cylinders, such that in the energy range
corresponding the fundamental bandgap only one waveguide mode can
propagate. Band diagram of the waveguide mode is shown in
Fig.\ref{fig5_band_str} (a).

\begin{figure}[tbp]
\begin{center}
\includegraphics[scale = 0.4]{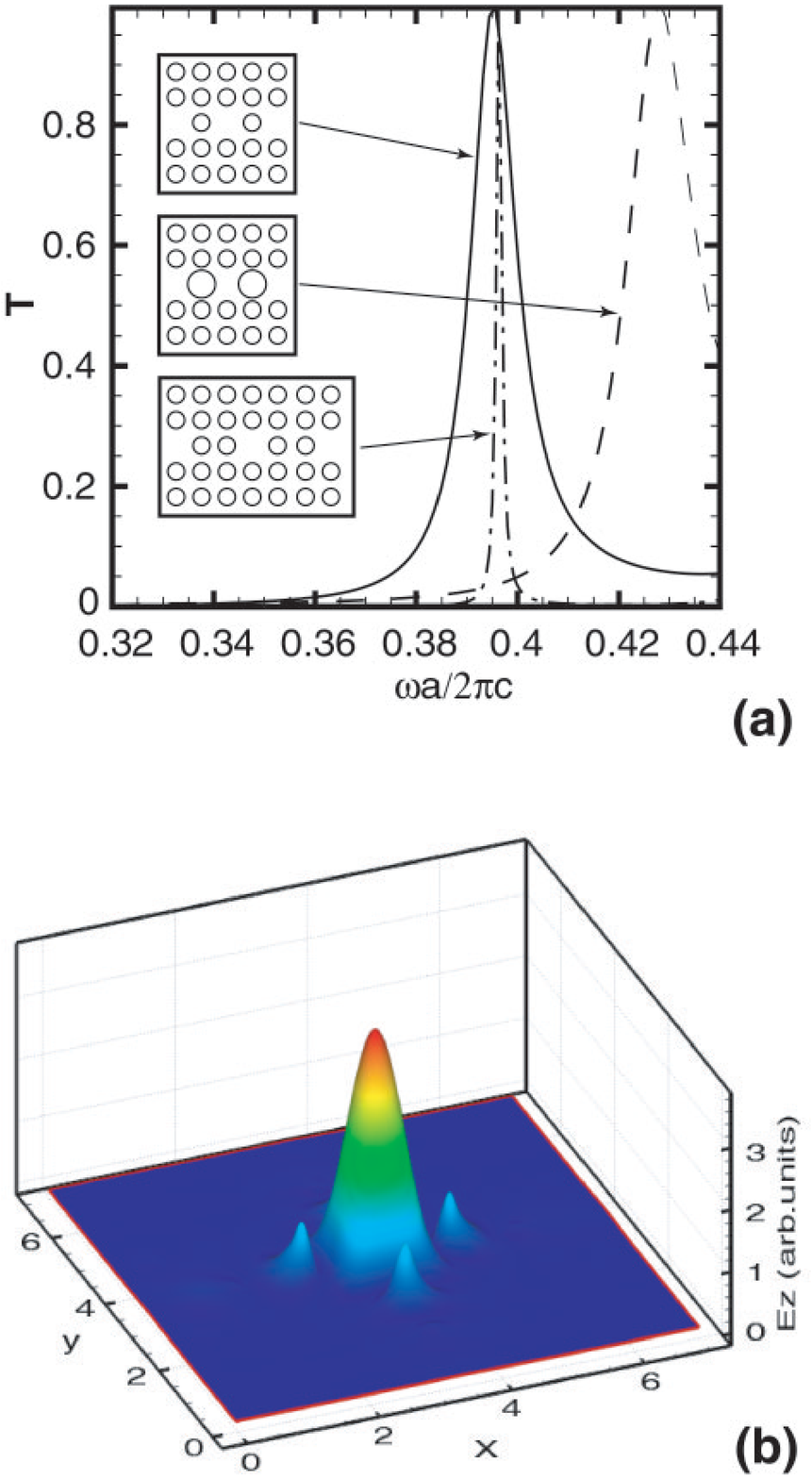}
\end{center}
\caption{(color online) (a) Transmission coefficient of three cavity
structures versus frequency. (b) Intensity of the $E_{z}$-component of the
electromagnetic field in the double-wall cavity at the resonance ($\protect%
\omega a / 2\protect\pi c =0.3952$). }
\label{fig7_trans_coeff}
\end{figure}

Three different cavities are introduced in order to show the
effect of geometry and demonstrate the importance of proper design
of a cavity. The first cavity is defined by two rods placed on the
lattice sites, see insets in Fig.\ref{fig7_trans_coeff}. In the
second structure the diameter of the rods is doubled, and for the
third cavity we place two rods from each side of the cavity to
achieve better confinement. A dependence of the transmission
coefficient on the incoming wave frequency is depicted in Fig.\ref%
{fig7_trans_coeff}(a). We would like to stress that in the calculation of
the transmission coefficient, the incoming, transmitted and reflected states
are the Bloch states of a waveguide (shown in Fig.\ref{fig5_band_str}(c)),
such that all spurious reflections from PC interfaces or computational
domain boundaries are avoided.

The fundamental parameter of cavity resonances is their $Q$-factor defined
as $Q$=2$\pi \omega $*(stored energy)/(energy lost per cycle), which can be
rewritten in the following form:
\begin{equation*}  \label{quality_factor}
Q=\omega \frac{\Omega }{4\int S_{in}dy}
\end{equation*}%
where $\Omega _{TM}=\int [\varepsilon \varepsilon _{0}|E_{z}|^{2}+\mu
_{0}(|H_{x}|^{2}+|H_{y}|^{2})]dxdy$\ \ and $\Omega _{TE}=\int [\mu
_{0}|H_{z}|^{2}+\varepsilon \varepsilon _{0}(|E_{x}|^{2}+|E_{y}|^{2})]dxdy$\
characterizes the energy stored in the system respectively for TM and TE
polarizations and the integral over $S_{in}$ is the incoming energy flux.
Eq. (\ref{quality_factor}) can be also expressed as a well-known relation $%
Q=\omega /\Delta \omega $ where $\omega $ is the resonant frequency and $%
\Delta \omega $ is the width of the resonant peak at half-maximum.

The resonance peak for the single-wall cavity is centered at
$\omega a / 2\pi c =0.3952$ and has $Q$ factor 35.5. As expected,
the highest $Q$ factor (327.7) is achieved for the case of
double-rod walls. Resonance peak in the case of larger rods is
shifted to the higher energy values ($\omega a / 2\pi c =0.4281$)
because of the decrease of the effective size of the cavity. The
lower $Q$ factor in this case (25.07) is because the larger rods
disrupt destructive interference in a bandgap of the PC.

Note that the width of the supercell used in the computations has
to be large enough to ensure that the intensity of the field
decays to zero at the domain boundaries. At the same time, it is
desirable to have the size of the computational domain as small as
possible. For the present computations, keeping this trade-off in
mind, we have chosen a supercell consisting of 7 unit cells in the
$y$-direction. This choice seems to be sufficient, as the field
intensity decreases by 5 orders of magnitude within the length of
two lattice constants from the waveguide towards the supercell
boundaries.

Finally, to confirm our results and to verify the developed method, we
performed calculations for the cavities and waveguides in PC studied by Li
\textit{et al.}\cite{Li2} and found a full agreement with their results.

\subsection{Surface states}

\begin{figure*}
\begin{center}
\includegraphics[
scale = 0.75 ]{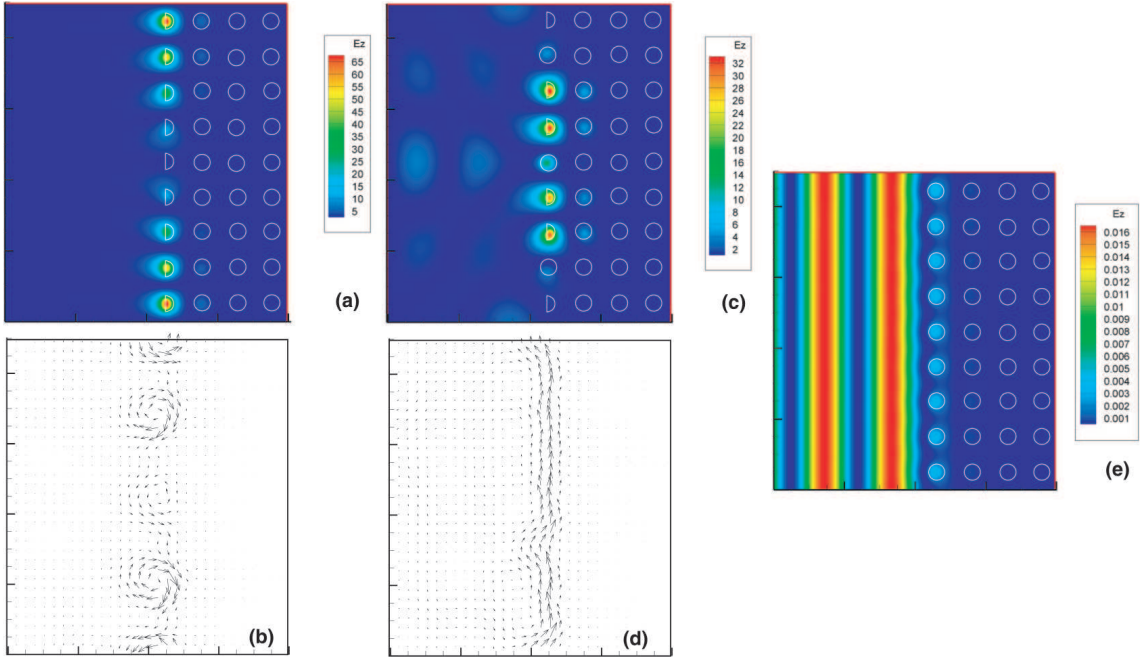}
\end{center}
\caption{(color online) $E_{z}$ field and Poynting vector
distributions for the structure 1 ((a),(b)) and for the structure
2 ((c),(d)) at the resonant
frequencies (marked by arrows in Fig. \protect\ref{fig_9_q_factor}). (e) $%
E_{z}$ field distributions for the structure that does not support surface
modes (a semi-infinite photonic crystal with all identical cylindrical
rods). In all cases the structures are illuminated by the incoming wave with
the incidence angle $\protect\alpha =\arctan {k_{y}/k_{x}=34.7}^{\circ }$.}
\label{fig_8_surf_fields}
\end{figure*}

In the previous section we considered wave propagation in an
infinite photonic crystal. Another aspect of interest is the
effect of the surface in semi-infinite photonic crystals that can
accommodate a localized state (surface mode) decaying both into
air and into a space occupied by the photonic
crystal\cite{JP,Elson_OPTE_2004}. In the present section we study
the coupling between an incident radiation and the surface states.
Note that a surface mode residing on the surface of an infinite
(in the $y$-direction) photonic crystal represents a truly bound
state with the infinite lifetime. However, because of the used
cyclic boundary conditions, our system is effectively confined in
the transverse direction. As the result, the translation symmetry
is broken, and the surface mode turns into a resonant
state with a finite lifetime. Using the developed method, we calculate the $%
Q $ factor of the surface modes. Our findings indicate that the surface
modes, thanks to their high $Q$ factors, can be used for lasing and sensing
applications.

We study two semi-infinite photonic crystal structures that
support localized surface modes. In the first case a surface row
of cylinders is composed of half-truncated rods\cite{JP}
(structure 1), and in the second case the cylindrical and
half-truncated rods in the surface row are interchanged as shown
in Fig. \ref{fig_8_surf_fields} (structure 2). In order to
calculate the $Q$ factor of the structures at hand, we illuminate
the semi-infinite photonic crystal by an incidence wave (that
excites the surface modes) and compute the intensity of the field
distribution. Note that the calculated field distribution includes
the contributions from both the surface mode exited by the
incident light, as well as the incident and reflected waves. This
leads to a nearly constant off-resonance background in the
dependence $Q=Q(\omega)$ that is caused by the contribution of the
incident and reflected waves in the total field intensity in Eq. (\ref%
{quality_factor}). To remove this background we calculate the $Q$
factor of a structure without surface states. We choose this
structure as a semi-infinite photonic crystal with all identical
cylindrical rods, which is known not to support surface modes
\cite{JP}. Then the obtained value is subtracted from the
calculated value of the $Q$ factor of the system under study. Note
that in the calculation of the $Q$ factor, the surface integration
in Eq. (\ref{quality_factor}) is performed over the area depicted
in Fig. \ref{fig_8_surf_fields}.

\begin{figure}
\begin{center}
\includegraphics[
scale = 0.45 ]{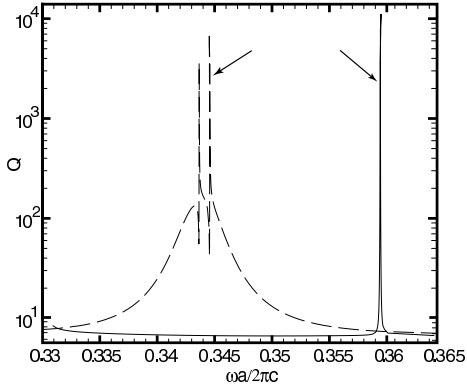}
\end{center}
\caption{Dependencies $Q=Q(\protect\omega)$ for structures 1 and 2
(solid and dashed lines respectively). Arrows indicate the
resonances for which the
field intensities and Poynting vectors are visualized in Fig. \protect\ref%
{fig_8_surf_fields}}
\label{fig_9_q_factor}
\end{figure}

Figure \ref{fig_9_q_factor} shows a $Q$ factor of structures 1 and
2 as a function of the frequency of the illuminating light. For
both structures the $Q$ factor reaches $\sim 10^{4}$. Figures
\ref{fig_8_surf_fields} (a),(c) show $E_{z}$-field distribution
for structures 1 and 2 at the resonance. For a comparison, a field
distribution for a structure that does not support a surface mode
(a semi-infinite photonic crystal with all identical cylindrical
rods) is shown in Figure \ref{fig_8_surf_fields} (e). In the
latter case the field intensity rapidly decays into the bulk of
the photonic crystal, whereas for the structures supporting the
surface modes, the intensity is strongly localized at the boundary
row of rods. It is also worth to mention that for the latter case
the intensity of the field in the surface mode exceeds the
incoming light intensity by 4 orders of magnitude, such that the
light intensity in the air region is not visible in the figures (compare \ref%
{fig_8_surf_fields} (a),(c) with (e)).

One can easily estimate the position of the resonant frequency for
the surface modes. Indeed, the outermost row of the cylinders
(where the surface state resides) can be considered as a resonator
with the characteristic resonant wavelengths following from the
cyclic boundary conditions and given by $\lambda _{\alpha
}=2\pi/k_{\alpha }$, where
\begin{equation}
 k_{\alpha
}=\frac{2\pi \alpha }{w},  \label{resonant_frequency}
\end{equation}%
$\alpha $ is the mode number and $w$ is the width of the
waveguide. The surface state for structure 1 exists only in a
limited frequency interval, $0.33\lesssim \omega a /2\pi c\lesssim
0.37$ (the dispersion relation of the surface mode of this
structure is given in Ref. \onlinecite{JP}). It follows from this
dispersion relation that all the modes given by Eq.
(\ref{resonant_frequency}), except $\alpha =4,$ are situated
outside this interval, whereas  the mode $\alpha =4$ corresponds
to the frequency $\omega a/2\pi c=0.365 $. This estimated
frequency agrees very well with the actual calculated
resonant frequency $\omega a/2\pi c\approx 0.359$, see Fig. \ref%
{fig_9_q_factor}. Note that the calculated field distribution, Fig. \ref%
{fig_8_surf_fields} (a), is fully consistent with the expected
pattern for 4th mode in the system of $n=9$ cylinders. (This field
distribution is determined by the overlap of the eigenstate
corresponding to the eigenfrequency (\ref{resonant_frequency})
with the actual positions of the cylinders in the outermost row).
Note that we performed calculations for different numbers of
cylinders in the transverse directions ($n=5,...,11$), and we
always find an excellent agreement with the predicted value of the
resonant frequency $\omega _{\alpha }.$

Figures \ref{fig_8_surf_fields} (b),(d) show Poynting vector distribution
for both structures at the resonance. For the structure 1 the Poynting
vector is ``curling" along the boundary, showing a low speed of the surface
state. In contrast, for the structure 2, the Poynting vector exhibits a
rapid flow of energy along the boundary. Another difference between these
structures is a very broad and rather strong ``background" peak in the
structure 2 in the region $0.34 \lesssim \omega a/2\pi c \lesssim 0.35$
(with $Q$ factor up to $\sim 100$). The presence of such the peak indicates
that the corresponding surface state can be rather robust to various kinds
of imperfections that are always present in real structures and which are
known to broaden the resonances and lead to decrease of the $Q$ factor \cite%
{Q}. These two examples of photonic crystals illustrate, that with
proper structure design one can engineer and tailor properties of
the surface states into the required needs.

High values of the $Q$ factors of the surface modes residing at
the interface of the photonic crystal structures indicate that
these systems can be used for lasing and sensing applications. The
lasing effect has been demonstrated for
different photonic crystal structures including band-gap defect mode lasers%
\cite{defect_mode}, distributed feedback lasers \cite{feedback}, and
bandedge lasers\cite{bandedge}. Utilization of the high-Q factor of the
surface modes represent a novel way to sustain lasing emission. To achieve
lasing effect careful design of the surface and surface mode engineering
should be performed and the developed method seems to be a suitable tool for
this purpose. A detailed study of the surface modes for various surface
terminations, their $Q$-values, and dispersion relations will be reported
elsewhere.

\subsection{Waveguide coupled to the open space}

\begin{figure}[tbh]
\begin{center}
\includegraphics[
scale = 0.38
]{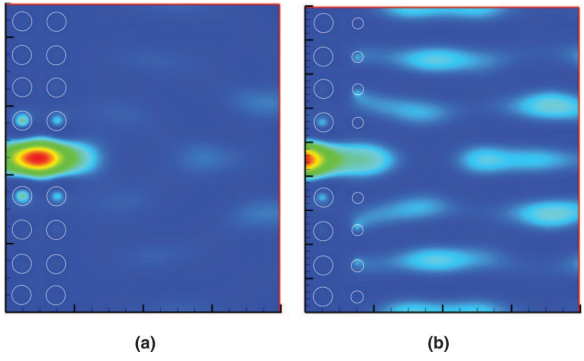}
\end{center}
\caption{(color online) $E_{z}$ field distributions at the surface of a
truncated photonic crystal with a waveguide. (a) The surface is composed of
cylinders with parameters identical to those in the bulk of the crystal, and
(b) the surface cylinders are two times smaller than the cylinders in the
bulk.}
\label{fig10_wg_opening}
\end{figure}

The last example of application of the method presented here is a
semi-infinite photonic crystal with a waveguide coupled to the surface, see
Fig. \ref{fig10_wg_opening}. It has been recently demonstrated that a
surface of a photonic crystal can serve as a kind of antenna to beam the
light emitted from the waveguide in a single direction\cite%
{Kramper_PRL_2004, Moreno_PRB_2004}. These findings outline the importance
of investigation of the surface modes in the photonic band-gap structures
that can eventually open up the possibilities to integrate such the devices
with conventional fiber optic devices.

In the present section we consider two different crystal terminations to
illustrate the effect of the surface on propagation of the light emitted
from the waveguide. In the first case the surface is composed of cylinders
with parameters identical to those in the bulk of the crystal, and in the
second case the surface cylinders are two times smaller than the cylinders
in the bulk.

The Bloch state propagating in a waveguide in the photonic crystal
couples with the states in air and the resulting field
distributions is shown in Fig. \ref{fig10_wg_opening}. The first
structure does not support the surface mode, and hence the light
intensity distribution in the air region exhibits a typical
diffraction pattern. However, for the case of the second structure
the field distribution in the air region is drastically different.
In this case the Bloch state in the waveguide couples with the
surface state localized at the crystal termination, such that the
whole surface acts as a source of radiation.

\section{Conclusions}

We have developed a method based on the recursive Green's function
technique for the numerical study of photonic crystal structures.
The method is proven to be an effective and numerically stable
tool for design and simulation of both infinite photonic crystals
and photonic crystals with boundaries. In the present method the
Green's function of the photonic structure is calculated
\textit{recursively} by adding slice by slice on the basis of
Dyson's equation. In order to account for the infinite extension
of the structure both into air and into the space occupied by the
photonic crystal we make use of the so-called \textquotedblleft
surface Green's functions" that propagate the electromagnetic
fields into infinity. This eliminates the spurious solutions
(often present in the conventional FDTD methods) related to e.g.
waves reflected from the boundaries defining the computational
domain. The developed method has been applied to scattering and
propagation of electromagnetic waves in photonic band-gap
structures including cavities and waveguides. In particularly, we
have shown that coupling of the surface states with incoming
radiation may result in enhanced intensity of the electromagnetic
field on the termination of the photonic crystal and very high
$Q$-factor of the surface modes localized at this termination.
This effect can be employed as an operational principle for
surface-mode lasers and sensors.

\acknowledgments A partial financial support from the National
Graduate School in Scientific Computing (A. I. R.) is
acknowledged.

\appendix

\section{Calculation of the transmission coefficient}

In this appendix we provide a detailed derivation of Eqs. (\ref{T}), (\ref{R}%
). Consider first an infinite periodic structure in a waveguide (supersell) geometry.
 The $\alpha $-th Bloch state in the lattice can be written in the form

\begin{equation}
|\psi _{\alpha }\rangle =\sum_{m,n}e^{ik_{\alpha }^{+}m}\phi _{m,n}^{\alpha
}\,a_{m,n}^{+}|0\rangle ,  \label{Bloch_3}
\end{equation}%
where summation is performed over all lattice sites and the function $\phi
_{m,n}^{\alpha }$ satisfies the conditions (\ref{Bloch}). Substituting Eq. (%
\ref{Bloch_3}) into Eq. (\ref{L_operator_2}), we arrive to the finite
difference equation valid for all sites $m,n$
\begin{gather}
v_{m,n}\phi _{m,n}^{\alpha }-u_{m,m+1;n,n}e^{ik_{\alpha}^+}\phi
_{m+1,n}^{\alpha }-u_{m,m-1;n,n}e^{-ik_{\alpha}^+}\phi _{m-1,n}^{\alpha }-
\label{disp} \\
-u_{m,m;n,n+1}\phi _{m,n+1}^{\alpha }-u_{m,m;n,n-1}\phi _{m,n-1}^{\alpha
}=\left( \frac{\omega \Delta}{c}\right) ^{2}\phi _{m,n}^{\alpha }.  \notag
\end{gather}%
Consider now the incoming state $|\psi _{\alpha }^{\mathrm{i}}\rangle ,$ Eq.
(\ref{s_Bloch}). Substituting Eq. (\ref{s_Bloch}) into Eq. (\ref%
{L_operator_2}) and using Eq. (\ref{disp}) we obtain%
\begin{align}  \label{aux1}
\left( \widehat{\mathcal{L}}-\left( \frac{\omega \Delta}{c}\right)
^{2}\right) |\psi_{\alpha}\rangle & = e^{ik_{\alpha
}^+}\sum_{n}u_{0,1;n,n}\phi _{1,n}^{\alpha }a_{0,n}^{+}|0\rangle - \\
& -\sum_{n}u_{1,0;n,n}\phi _{0,n}^{\alpha }a_{1,n}^{+}|0\rangle .  \notag
\end{align}%
Substituting this equation into Eq. (\ref{respons}), calculating the matrix
elements $\langle M+1,n|\psi \rangle $ and $\langle 0,n|\psi \rangle ,$ and
using the relations%
\begin{align}
G^{M+1,0}& =-G^{M+1,1}U_{1,0}\Gamma _{l}, \\
G^{0,0}& =\Gamma _{l}-G^{0,1}U_{1,0}\Gamma _{l},
\end{align}%
that follow from Dyson's equation, we arrive to Eqs. (\ref{T}),(\ref{R})
determinig the transmission and reflection amplitudes.

\section{Derivation of the Dyson's equation}

Let $\widehat{\mathcal{L}}^{0}$ be the operator describing an
unperturbed system and $\widehat{\mathcal{V\,}}$ be a
perturbation. In our case the unperturbed system consists of
several subsystems, e.g. $m$ slices of the internal structure and
$(m+1)$th slice, and the perturbation corresponds to the coupling
(hopping) between them (see Fig. \ref{fig2_Dyson}). The operator
of the total
(perturbed) system reads%
\begin{equation}
\widehat{\mathcal{L}}=\widehat{\mathcal{L}}^{0}+\widehat{\mathcal{V\,}}.
\label{perturbed_hamiltonian}
\end{equation}%
Let $G^{0}$ and $G$ be the Green's functions of the unperturbed
and the total (perturbed) systems respectively. Starting with the
definition of the Green's function
(\ref{Greens function}), we obtain%
\begin{equation}
G^{-1}=(\omega \Delta /c)^{2}-\widehat{\mathcal{L}}=(\omega \Delta /c)^{2}-%
\widehat{\mathcal{L}}^{0}-\widehat{\mathcal{V\,}}=(G^{0})^{-1}-\widehat{%
\mathcal{V\,}}.  \label{green_derivation1}
\end{equation}%
Multiplying this expression from the left with $G$ and from the right with $%
G^{0} $ we arrive to Dyson's equations

\begin{align}
G^{0}G^{-1}G& =G^{0}(G^{0})^{-1}G-G^{0}\widehat{\mathcal{V\,}}G\Rightarrow
\notag  \label{dyson_eq1} \\
G& =G^{0}+G^{0}\widehat{\mathcal{V\,}}G.
\end{align}%
Similarly one can also show that
\begin{equation}
G=G^{0}+G\widehat{\mathcal{V\,}}G^{0}.  \label{dyson_eq2}
\end{equation}

\section{The Green's function for a single slice}

\bigskip The operator describing the $m$-th slice has the form

\begin{align}
~\widehat{l_{m}}& =\sum_{n=1}^{N}(v_{m,n}a_{m,n}^{+}a_{m,n}-
\label{single slice} \\
& -u_{m,m;n,n+1}a_{m,n}^{+}a_{m,n+1}-u_{m,m;n+1,n}a_{m,n+1}^{+}a_{m,n}).
\notag
\end{align}%
Using this operator in the definition of Green's function (\ref{Greens
function}), and calculating the matrix elements $(...)_{m,m;n,n^{\prime
}}\equiv \langle 0|a_{m,n}...\,a_{m,n^{\prime }}^{+}|0\rangle ,$ we arrive
to the $N\times N$ system of linear equations for the matrix elements of the
Green's function of a single slice $g_{m},$%
\begin{equation}
\sum_{n^{\prime \prime }=1}^{N}\left( \left( \frac{\omega \Delta}{c}\right)
^{2}\delta _{n,n^{\prime \prime }}-l_{m,m;n,n^{\prime \prime }}\right)
g_{m,m;n^{\prime \prime },n^{\prime }}=\delta _{n,n^{\prime }},
\label{Eq:single slice}
\end{equation}%
where the matrix element of the operator $\widehat{l}_{m}$ reads,%
\begin{align}
& l_{m,m;n,n^{\prime \prime }}=v_{m,n}\delta _{n,n^{\prime \prime }}- \\
& -u_{m,m;n^{\prime \prime }-1,n^{\prime \prime }}\delta _{n,n^{\prime
\prime }-1}-u_{m,m;n^{\prime \prime }+1,n^{\prime \prime }}\delta
_{n,n^{\prime \prime }+1}.
\end{align}%
Note that because of the cyclic boundary conditions in the $n$-direction,
the matrix elements $u_{m,m;1,N}$ and $u_{m,m;N,1}$ are distinct from zero
and defined according to $u_{m,m;N,1}=u_{m,m;0,1},$ and $%
u_{m,m;1,N}=u_{m,m;N+1,N}.$\newline

\end{document}